# Macho Parallaxes From A Single Satellite

Andrew Gould

Dept of Astronomy, Ohio State University, Columbus, OH 43210

e-mail gould@payne.mps.ohio-state.edu

## Abstract

Massive Compact Objects (Machos) are currently being discovered at substantially higher rates than would be expected from standard models of known stellar populations. To determine whether they are due to non-standard distributions of known populations or to a heretofore unknown ('dark') population, one must acquire more information about the individual events. Space-based parallaxes are potentially the best tool for extracting additional information, yielding the "reduced" speed $\tilde v = v/(1-z)$, the "reduced" Einstein radius $\tilde r_e = [4GMD_{\rm OS}z/(1-z)]^{1/2}/c$, and the transverse direction of motion $\Phi$. Here $M$ is the Macho mass, $v$ is its transverse speed, and $z$ is its distance relative to the source distance, $D_{\rm OS}$. For example bulge and disk Machos can be cleanly separated by measuring $\tilde v$. To leading order, parallax measurements by a single satellite result in a four-fold degeneracy: two possible values of $\tilde v$ and two possible signs for the component of motion perpendicular to the projected satellite-Earth vector. It had been believed that a second satellite would be required to break this degeneracy. I show that the velocity difference between the satellite and the Earth allows one to partially or totally break the degeneracy using a single satellite. For most Macho events it is possible to measure $\tilde v$ and $\tilde r_e$. For some it is also possible to measure $\Phi$. The proposed Space Infrared Telescope could measure $\sim 100$ parallaxes per year by applying $\sim 400\,\rm hr$ of telescope time.

Subject Headings: astrometry – dark matter – gravitational lensing

## 1. Introduction

Several dozen candidate lensing events have now been detected toward the galactic bulge by the OGLE (Udalski et al. 1994 and references therein) and MACHO (Alcock et al. 1994; C. Stubbs, private communication 1994) collaborations. Estimates of the observed optical depth to lensing in this direction of $\tau \sim 3\text{--}4 \times 10^{-6}$ far exceed the sum of the values expected for a standard disk, $\tau_d \sim 5 \times 10^{-7}$ (Paczyński 1991; Griest et al. 1991) and a standard axisymmetric bulge, $\tau_b \sim 5 \times 10^{-7}$ (Kiraga & Paczyński 1994). Whether the additional observed events are due to additional material in the disk, an elongated bulge, a highly flattened halo, or to some other structure, they are potentially very exciting. Extrapolating from current detection rates, I estimate that the MACHO collaboration could see $\sim 100$ events next bulge season if they monitor 35 fields.

The MACHO (Alcock et al. 1993; K. Griest, private communication 1994) and EROS (Aubourg et al. 1993) collaborations have seen a total of 5 candidate events toward the Large Magellanic Cloud (LMC). Although far fewer than expected from a standard spherical halo, this is also far more than would be expected from a combination of the known stars in the disk, thick disk, and spheroid (Gould, Miralda-Escudé, & Bahcall 1994; Bahcall et al. 1994) or from the LMC itself (Gould 1994c). Thus the source of these events is also a puzzle.

In the course of detecting a Massive Compact Object (Macho) by means of its time-varying magnification of a source star, one generally obtains only one piece of information about the Macho itself, the time scale $\omega^{-1}$. This is a combination of three of the four physical parameters that one would like to know:

$$\omega^{-1} = \frac{[4GMD_{\rm OS} z(1-z)]^{1/2}}{vc}, \tag{1.1}$$

where $M$ is the Macho mass, $v$ is its transverse speed relative to the observer-source line of sight, and $z$ is the ratio of its distance $D_{\rm OL}$ to that of the source $D_{\rm OS}$,

$$z \equiv \frac{D_{\rm OL}}{D_{\rm OS}}. \tag{1.2}$$

The source distance $D_{\rm OS}$ is assumed to be known from spectroscopic parallax. The fourth potentially measurable parameter is $\Phi$, the direction of transverse motion. The direction does not enter the time scale $\omega^{-1}$. The distribution of observed time scales is generally the only information available about the Machos. Even when this is measured over several lines of sight, the implications of such measurements are highly degenerate, especially because we are certainly seeing two distinct Macho distributions and very probably more than two. Hence, in order to understand the nature of the events it is necessary to extract more information about them.

Space-based acquisition of Macho parallaxes is potentially the most powerful method of obtaining additional information (Gould 1992; Gould 1994b, hereafter Paper I). First, Macho parallaxes yield two additional pieces of information about each event. When combined with the time scale, the three parameters are the "reduced" Einstein radius $\tilde{r}_e$, the "reduced" transverse speed $\tilde{v}$,

$$\tilde{r}_e^2 \equiv \frac{4GD_{\rm OS}}{c^2} \frac{Mz}{1-z}, \qquad \tilde{v} \equiv \frac{v}{1-z}, \qquad (1.3)$$

and $\Phi$, the transverse direction relative to the satellite-Earth separation vector. For example, the reduced speed can discriminate well between disk and bulge Machos seen toward the bulge, or between Galactic and LMC objects seen toward the LMC (Han & Gould, in preparation). Once these populations are distinguished, the reduced transverse speed and direction give direct kinematic information about each population, and the reduced Einstein ring gives information about the mass spectrum. A second important feature is that once the appropriate satellite is launched, space-based parallaxes can be obtained for a large fraction of all events. By contrast, special circumstances are required to obtain ground-based parallaxes (Gould et al. 1994) or proper motions (Gould 1994a, Nemiroff & Wickramasinghe 1994; Maoz & Gould 1994).

Parallaxes are determined by measuring the difference in the times of maximum magnification $\Delta t$, and the difference in the impact parameters $\Delta\beta$, as determined from the Macho light curves seen from the Earth and a satellite. See Figures 1 and

2 of Paper I. The light curves are described by the magnification $A[x(t)]$,

$$A(x) = \frac{(x^2 + 2)}{x(x^2 + 4)^{1/2}}, \qquad x(t) = [\omega^2(t - t_0)^2 + \beta^2]^{1/2}, \qquad (1.4)$$

where $t_0$ is the time of maximum magnification and $\beta$ is the impact parameter in units of the Einstein radius, $r_e = [4GMD_{\mathrm{OS}}z(1-z)]^{1/2}/c$. The vector displacement in the Einstein ring

$$\Delta \mathbf{x} \equiv (\omega \Delta t, \Delta \beta), \qquad (1.5)$$

is related to the two-dimensional separation vector $\mathbf{r}$ between the satellite and the Earth by,

$$\Delta \mathbf{x} = \frac{\mathbf{r}}{\tilde{r}_e}. \qquad (1.6)$$

Unfortunately, while $\omega \Delta t = \omega(t'_0 - t_0)$ is well defined by comparing the value of $t'_0$ measured from the satellite light curve with $t_0$ measured from Earth, the impact-parameter difference $\Delta \beta$ has a four fold degeneracy: $\Delta \beta$ can have two distinct magnitudes,

$$\Delta \beta_\pm \equiv |\beta' \pm \beta|, \qquad (1.7)$$

corresponding to impact parameters on the opposite or same side of the source. These different values for $\Delta \beta$ lead to corresponding different values for $\Delta x_\pm = [(\omega \Delta t)^2 + (\Delta \beta_\pm)^2]^{1/2}$ and so for $\tilde{v} = \omega r/\Delta x_\pm$. In addition, $\Delta \beta$ can be of either sign. The two signs correspond to the two directions perpendicular to the projected satellite-Earth separation $\mathbf{r}$, i.e., $\Phi = \pm \tan^{-1}(\Delta \beta/\omega \Delta t)$. In Paper I, I discussed the possibility of breaking this four-fold degeneracy by making observations from a second satellite. However, since the launching of even one satellite into solar orbit is a formidable undertaking, the need for a second satellite would be a significant drawback.

Here I show that the degeneracy in $\Delta\beta$ can be broken by measuring the difference in time scales

$$\Delta\omega = \omega' - \omega, \tag{1.8}$$

between the satellite and the Earth.

## 2. Breaking the Degeneracy

Let the three-dimensional separation vector between the satellite and the Earth be denoted $\mathbf{R}$, and let their relative velocity be denoted $\mathbf{U}$. Let $\hat{\mathbf{s}}$ be the direction of the source. Then the projected separation $\mathbf{r}$ and projected velocity $\mathbf{u}$ are given by

$$\mathbf{r} = \mathbf{R} - (\mathbf{R} \cdot \hat{\mathbf{s}})\hat{\mathbf{s}}, \qquad \mathbf{u} = \mathbf{U} - (\mathbf{U} \cdot \hat{\mathbf{s}})\hat{\mathbf{s}}. \tag{2.1}$$

I define the parallel and perpendicular components of $\mathbf{u}$ by

$$u_\parallel \equiv \frac{\mathbf{r} \cdot \mathbf{u}}{r}, \qquad u_\perp \equiv \frac{(\mathbf{r} \times \mathbf{u}) \cdot \hat{\mathbf{s}}}{r}. \tag{2.2}$$

I assume that the projected velocity difference is small compared to the reduced speed $u \ll \tilde{v}$. The difference in inverse time scales $\Delta\omega$ is then given by

$$\frac{\Delta\omega}{\omega} = \frac{|\tilde{\mathbf{v}} + \mathbf{u}| - \tilde{v}}{\tilde{v}} \simeq \frac{\tilde{v}_\parallel u_\parallel + \tilde{v}_\perp u_\perp}{\tilde{v}^2}. \tag{2.3}$$

Since $\tilde{v} = \omega \tilde{r}_e$, equation (1.6) (or eq. 2.8 from Paper I) implies,

$$\tilde{v}_\parallel = \frac{r\omega^2 \Delta t}{(\Delta x)^2}, \qquad \tilde{v}_\perp = \frac{r\omega \Delta\beta}{(\Delta x)^2}, \qquad \tilde{v} = \frac{r\omega}{\Delta x}. \tag{2.4}$$

Thus, equation (2.3) becomes

$$u_\perp \Delta\beta = r\Delta\omega - u_\parallel \omega \Delta t. \tag{2.5}$$

Since $\mathbf{u}$ and $\mathbf{r}$ are known from the satellite's orbit and $\Delta\omega$ and $\omega\Delta t$ are known from the comparison of light curves, $\Delta\beta$ is unambiguously determined by equation (2.5). The only question is: is $\Delta\beta$ well enough determined to break the four-fold degeneracy which arises in the previous analysis?

For definiteness I will assume that $\delta_\omega$, the error in estimating $\Delta\omega/\omega$ from the two light curves is given by

$$\delta_\omega = 0.01, \tag{2.6}$$

and that the errors in estimating $\omega\Delta t$, $\Delta\beta_-$, and $\Delta\beta_+$ are respectively

$$\delta_t = 0.7\delta_\omega, \qquad \delta_- = 0.5\delta_\omega, \qquad \delta_+ = 8\delta_\omega. \tag{2.7}$$

In § 6, I will discuss the reasons for the relative error sizes given by equation (2.7) and the prospects for achieving the accuracy given by equation (2.6).

If information about $\Delta\omega$ is ignored, then the $\chi^2$ surface over the $\Delta\mathbf{x} = (\omega\Delta t, \Delta\beta)$ plane will have four minima. The minima at $(\omega\Delta t, \pm\Delta\beta_-)$ will have $1\,\sigma$ error ellipses that are $0.007 \times 0.005$ and the minima at $(\omega\Delta t, \pm\Delta\beta_+)$ will have $1\,\sigma$ error ellipses that are $0.007 \times 0.08$. To break the degeneracy, the information encoded in $\Delta\omega$ must distinguish between these minima at the $\sim 3\,\sigma$ level.

## 3. Satellite With An Earth-Like Orbit

To address the problem of degeneracy breaking concretely, I assume that the satellite is in an Earth-like orbit, but displaced from the Earth by an angle $\theta$. Then $R = 2\sin(\theta/2)R_\oplus$, $U = 2\sin(\theta/2)v_\oplus$, and $\mathbf{R}\cdot\mathbf{U} = 0$. Here $v_\oplus = 30\,\mathrm{km\,s^{-1}}$ is the Earth's orbital speed and $R_\oplus = 1\,\mathrm{AU}$. Let $\alpha$ be the angle between $\hat{\mathbf{s}}$ and the south ecliptic pole, and let $\psi$ be the phase of the orbit such that at $\psi = 0$, the projected displacement vector $\mathbf{r}$ is most closely aligned with $\hat{\mathbf{s}}$. I then find

$$\begin{aligned} r &= R(1 - \sin^2\alpha\cos^2\psi)^{1/2}, & u &= \Omega_\oplus R(1 - \sin^2\alpha\sin^2\psi)^{1/2}, \\ u_\perp &= -\Omega_\oplus R\frac{R}{r}\cos\alpha, & u_\parallel &= \Omega_\oplus R\frac{R}{r}\sin^2\alpha\frac{\sin 2\psi}{2}, \end{aligned} \tag{3.1}$$

where $\Omega_\oplus \equiv v_\oplus/R_\oplus = 2\pi\,\mathrm{yr}^{-1}$.

## 4. Observations Toward the LMC

Since the LMC is very nearly at the south ecliptic pole, $\alpha \sim 0$, equation (3.1) implies that $r = R$, $u = U$, and

$$u_\perp = -R\Omega_\oplus, \qquad u_\| = 0. \tag{4.1}$$

Hence, equation (2.5) becomes

$$\Delta\beta = -\frac{\Delta\omega}{\Omega_\oplus}. \tag{4.2}$$

The smallest separation in $\Delta\beta$ between the four solutions will generally be $\mathcal{O}(\Delta x)$. In any event, this is the smallest separation that needs to be resolved since if two solutions are much closer than this they represent essentially the same reduced velocity. Combining equations (2.4), (2.6) and (4.2), I find the condition for complete degeneracy breaking: $u_\perp \gtrsim 3\delta_\omega \tilde{v}$ or

$$\frac{v_\oplus}{\tilde{v}}\frac{R}{R_\oplus} \gtrsim 0.03 \qquad \text{(vector solution)}. \tag{4.3}$$

Note that for Machos in the Galactic halo, $\tilde{v}/v_\oplus \sim 10$, while for LMC Machos $\tilde{v}/v_\oplus \gtrsim 50$. From equation (4.3), it might appear that the best strategy is to place the satellite near $\theta \sim 180°$ so that $R \sim 2R_\oplus$. In fact, one must be careful that $R \lesssim \tilde{r}_e$ since otherwise the event as seen from the satellite will be out of the Einstein ring. Equation (4.3) shows that the degeneracy can be broken for Galactic Machos provided that $R \gtrsim 0.3\,R_\oplus$ but cannot be broken for LMC Machos unless $R > R_\oplus$ or the accuracy is improved beyond equation (2.6). It is much easier, however, to distinguish between the $\pm\Delta\beta_-$ solutions from the $\pm\Delta\beta_+$ solutions since these are separated by $\mathcal{O}(1)$ rather than $\mathcal{O}(\Delta x)$. From equations (2.6) and (4.2) I find the less restrictive condition: $\Omega_\oplus \gtrsim 3\delta_\omega \omega$ or

$$\omega^{-1} \gtrsim 1.7\,\text{days} \qquad \text{(scalar solution)}. \tag{4.4}$$

When equation (4.4) holds, $\tilde{v}$ and $\tilde{r}_e$ can be measured, but $\Phi$ cannot be measured unless equation (4.3) also holds.

## 5. Observations Toward the Bulge

Observations toward the bulge cover many lines of sight. For definiteness, I adopt a line of sight with an angle $\alpha = 84°$ from the south ecliptic pole, similar to Baade's Window. That is, $\cos\alpha = 0.10$.

The fact that the bulge lies near the ecliptic introduces new complications in the parallax measurement which are not present for the LMC. The size of the displacement in the Einstein ring $\Delta\mathbf{x} = (\omega\Delta t, \Delta\beta)$ between the Earth and the satellite is a strong function of the time of year. Since $\Delta x = r/\tilde{r}_e$, the measurement is most effective if the projected displacement is as large as possible without being larger than the reduced Einstein ring. From equation (3.1) it is clear that $r$ varies by a factor $\sec\alpha \sim 10$ over the course of a year. Thus, no matter what physical separation $R$ is chosen and no matter what is the distribution of $\tilde{r}_e$, the projected separation will be non-optimal during some parts of the year. Nevertheless, the difficulties presented by the changing length of $r$ are not in and of themselves very serious. Typical events arising from disk Machos have $\tilde{r}_e \sim 2\,\mathrm{AU}\,(M/0.1\,M_\odot)^{1/2}$ while those in the bulge have $\tilde{r}_e \sim 5\,\mathrm{AU}\,(M/0.1\,M_\odot)^{1/2}$. During $\sim 80\%$ of the year the projected separation is in the range $0.33 \le r/R \le 1$. Hence, if $R \sim R_\oplus$ then over most of the year and for most Machos, $0.04 \lesssim \Delta x \lesssim 1$, so that according to equation (2.7), it should be possible to measure $\Delta\mathbf{x}$ to within $\lesssim 20\%$ (up to a four-fold degeneracy).

The fact that the bulge lies near the ecliptic does pose significant problems for breaking the degeneracy. The error in estimating $\Delta\beta$ in the degeneracy-breaking equation (4.2) has two terms, $(r\omega/u_\perp)\delta_\omega$ and $(u_\parallel/u_\perp)\delta_t$. As I discuss below, the second term can effectively be ignored. Thus the condition for complete degeneracy breaking is $u_\perp \gtrsim 3\delta_\omega \tilde{v}$ or

$$\frac{v_\oplus}{\tilde{v}} \frac{R}{R_\oplus} (1 + \tan^2\alpha \sin^2\psi)^{-1/2} \gtrsim 0.03 \qquad \text{(vector solution)}. \qquad (5.1)$$

Note that this expression is smaller than the ecliptic-pole formula (4.3) by a factor $(1+\tan^2\alpha\sin^2\psi)^{-1/2}$. This factor remains in the range 0.1–0.2 for about 2/3 of the year. For disk Machos, typical reduced speeds are $\tilde{v} \sim 300\,\mathrm{km\,s^{-1}}$, implying that for most of the year few disk lensing events can be completely resolved. Since bulge Machos have substantially higher $\tilde{v}$, it will generally not be possible to completely resolve these events.

However, it will generally be possible to distinguish between the $\Delta\beta_+$ and $\Delta\beta_-$ solutions and thereby measure $\tilde{v}$ and $\tilde{r}_e$ for the majority of Machos presently being detected. Following the same argument that led to equation (4.4), I find $\Omega_\oplus \gtrsim 3\delta_\omega \omega$ or

$$\omega^{-1} \gtrsim 1.7(1 + \tan^2\alpha\sin^2\psi)^{1/2}\,\mathrm{days} \qquad \text{(scalar solution).} \qquad (5.2)$$

I now justify ignoring the error arising from the second term in equation (2.5). From equations (2.7) and (3.1), one finds that the ratio of the errors of the second relative to the first term is $0.35\,(\Omega_\oplus/\omega)(1 - \sin^2\alpha\cos^2\psi)^{-1}\sin(2\psi)$. This is greater than unity only if $\omega^{-1} > 17\,\mathrm{days}$. Even in this case, the second term is significant for only a short fraction of the event. Hence, it has no major effect.

## 6. Error Estimates and Systematic Effects

I derived the estimates of the error ratios (2.7) by assuming that a "typical" lensing event ($\beta \sim \beta' \sim 0.5$) is measured frequently and with uniform accuracy from both the ground and space over a time interval $3\omega^{-1}$, beginning when the source enters the Einstein ring. The absolute normalization [eq. (2.6)] was set by assuming that the photometry has errors of $\sigma = 0.01$ mag and is carried out with frequency $f = 10\omega$. For different parameters, the errors scale $\propto \sigma f^{-1/2}$.

A very important implicit assumption of the calculation is that the satellite and ground-based observations are carried out using very similar filters. The importance of this assumption can be seen as follows. Under the observational assumptions listed above, it is typically possible to measure the impact parameter $\beta$ of an individual light curve only to an accuracy of $\pm 0.12$. Similarly it is possible to measure $\omega$ only to a fractional accuracy of $\pm 18\%$. At first sight these high errors make the estimates for the uncertainties in $\Delta\beta_-$ and $\Delta\omega$ given in equations (2.6) and (2.7) appear ludicrous. However, the reason for the high errors in $\beta$ and $\omega$ is that one must allow for light from an additional unresolved star, either the lensing star, a companion to the lens or the source, or a random superposition of an unrelated star. If such an unresolved star is present and not taken into account, it will make the impact parameter appear lower and the event shorter. The fit to the light curve is very insensitive to the fraction of light that is not being lensed, and errors in this estimate induce errors in $\beta$ and $\omega$. However, if the light curves from the ground and space are both obtained with the same filter then one knows *a priori* that the fraction of the blended source "star" due to background light is the same for both sets of observations. Hence the differences $\Delta\beta_-$ and $\Delta\omega$ can be determined much more accurately than the individual parameters $\beta$ and $\omega$. On the other hand, the sum $\Delta\beta_+$ is rather poorly determined [see eq. (2.7)].

Let $\overline{\lambda}$ and $\overline{\lambda'}$ be the mean effective wavelengths of the Earth and satellite filters, and define $\Delta\lambda \equiv \overline{\lambda'} - \overline{\lambda}$. I then find that errors in $\Delta\omega/\omega$ as large as $\sim \Delta\lambda/\overline{\lambda}$ could result. Thus, to avoid compromising the accuracy of the experiment, one should insure $\Delta\lambda \lesssim 0.25\delta_\omega\overline{\lambda}$ or $\Delta\lambda \lesssim 20\text{\AA}$ in $I$ band. If this level of precision cannot be achieved, then the ground-based observations must be done in two bands in order to find the color correction term between the ground-based and space-based filters.

There are two systematic effects that could compromise the degeneracy-breaking measurement of $\Delta\omega$: either the source or the lens could be a binary. If the source star suffers acceleration due to a companion with component $a_\perp$ perpendicular to $\mathbf{r}$, the satellite and Earth light curves will differ very much as they would if $u_\perp = a_\perp \Delta t z/(1-z)$. For the great majority of binary sources, this value of $u_\perp$ will be clearly inconsistent with being due to the known satellite-Earth velocity difference. On the other hand some will be consistent. Hence one must worry that any measured $\Delta\omega$ that is consistent with an allowed value of $\Delta\beta$ [see eq. (2.5)], is actually produced by a companion. The inferred acceleration would be

$$a_\perp = a_\oplus \frac{v}{v_\oplus} \frac{\cos\alpha}{1 - \sin^2\alpha \cos^2\psi} \frac{1}{z\cos\Phi}, \tag{6.1}$$

where $a_\oplus$ is the Earth's acceleration. In most cases this acceleration would be so large that it could easily be detected by spectroscopic observations (provided that the orbit was not very close to the plane of the sky). All sources with non-zero measured $\Delta\omega$ should be monitored spectroscopically to check for possible duplicity.

In general, the effects produced by binary lenses do not resemble those from satellite-Earth velocity differences so that the contamination from this effect will be very small.

## 7. Satellite Requirements

The optimum value of $R$ is different for different populations of Machos. For Galactic Machos seen toward the LMC and disk Machos seen toward the bulge, the separation should be $R \lesssim R_\oplus$ so that the event as seen from the satellite does not fall outside the Einstein ring. For bulge Machos and LMC Machos, $R$ should be as big as possible to maximize the relatively small value of $\Delta x$. These divergent requirements can be partially reconciled by launching a satellite into solar orbit which stays in an Earth like orbit but gradually drifts away from the Earth. The satellite-Earth observations would then be most sensitive to nearby Machos at the beginning and more sensitive to distant Machos later on.

At present, it appears that $\sim 100$ Machos will be discovered toward the bulge during each $\sim 6$ month bulge season. To observe each event 30 times requires $\sim 16$ observations per day. If the typical star has $I \sim 19$, then 1% photometry could be obtained in 1.5 hr/day on a 1m telescope or in 24 hr on a 0.25m. Similar accuracy could be achieved in $K$ band assuming $(I - K)_0 \sim 0.5$ and $A_V \gtrsim 1.5$. A 1m telescope would have the advantage of better resolution and there are many other imaging projects one might want to do with such an instrument. In fact, the proposed Space Infrared Telescope Facility (SIRTF) could make very useful parallax observations with a commitment of only $\sim 400\,\mathrm{hr/yr}$. As presently designed, SIRTF would drift $0.1\,\mathrm{AU\,yr^{-1}}$ behind the Earth. To optimize the satellite for Macho observations, this rate should be about doubled.

**Acknowledgements**: I would like to thank D. DePoy for helping to clarify several points.